# VDMN: A GRAPHICAL NOTATION FOR MODELLING VALUE DRIVER TREES

**Benjamin Matthies**
Münster School of Business, FH Münster – University of Applied Sciences, Germany

**Abstract:** Value Driver Trees (VDTs) are conceptual models used to illustrate and analyse the causal relationships between key performance indicators and business outcomes, thereby supporting managerial decision-making and value-based management. Despite their increasing application, there are still no systematic guidelines for the modelling of such conceptual models. To fill this gap, this study introduces the Value Driver Modelling Notation (VDMN), a graphical notation developed to systematically guide VDT modelling. This notation includes a comprehensive set of semantic constructs and an intuitive graphical syntax. To evaluate its practical utility, the VDMN was applied in two case studies and assessed through expert interviews. The results show that the notation supports a consistent and comprehensible modelling of VDTs. The VDMN thus represents a significant step towards the systematisation and standardisation of VDT modelling.

## 1 Introduction

Value Driver Trees (VDTs) are an increasingly used tool for the systematic representation and analysis of value creation in organisations (Koller et al., 2020; Matthies, 2024; Demyttenaere et al., 2023). VDTs use a hierarchical structure to illustrate the causal relationships between operational activities, financial metrics and overall business performance. Based on this concept, the influence of the most important performance indicators – the so-called value drivers (e.g., 'conversion rate' or 'scrap rate') – on the subsequent financial indicators (e.g., 'sales' or 'production costs') can be modelled in order to calculate and simulate cause-and-effect relationships with regard to a key business indicator (e.g., 'profit'). VDTs therefore provide management with a sound basis for decision making, as they enable a simulation-based evaluation of alternative decision options and thus a structured prioritisation of actions in terms of their value contribution to the company (Koller et al., 2020). Therefore, the concept of VDTs has gained increasing attention in research (Matthies, 2024) and practice (Demyttenaere et al., 2023).

Despite the increasing prevalence and practical benefits of VDTs, there is an apparent lack of systematic and standardised approaches to modelling such indicator-based models in the scientific literature. An analysis of published VDTs shows great heterogeneity in terms of structure, detail, visualisation and semantic depth (Matthies, 2024). This lack of standardisation not only complicates the systematic development and interpretation of VDTs in practice, but also makes it difficult to integrate them efficiently into modern, data-driven decision support systems (Diana, 2021; Morgret et al., 2024; Wobst et al., 2023). In order to fully exploit the potential of VDTs – for applications such as management reporting (Sidler and Gerussi, 2023), predictive forecasting (Valjanow et al., 2019), AI-supported business analysis or simulation-based planning (Duckstein, 2023; Gupta et al., 2022; Feldmann and Matthies, 2025) – a formalised, conceptually sound graphical notation for modelling VDTs will therefore be beneficial.

The goal of this study is to address this gap by proposing and evaluating a graphical notation for modelling VDTs. The notation is based on a systematic classification of 34 semantic model constructs derived from an extensive literature review of 161 real VDTs using a taxonomy design process (see Matthies, 2024). Based on this classification, a visual syntax was defined that allows a structured, consistent and interpretable representation of typical VDTs. The research objective (RO) can therefore be summarised as follows:

*RO: Development and evaluation of a graphical notation for modelling VDTs.*

The proposed notation was evaluated on the basis of two practice-oriented case studies, in which exemplary VDTs were re-modelled using the proposed notation and evaluated by experts. The aim of the evaluation was to test the utility of the notation against key quality requirements for conceptual modelling languages (see, e.g., Moody, 2005, 2009; Paige et al., 2000; Burton-Jones et al., 2009; Wand and Weber, 1993). In this respect, the models were assessed not only as complete and correct, but also as comprehensible, logical and reliable in terms of interpretation. The proposed notation thus fulfils key requirements for the quality of conceptual modelling languages and represents a promising basis for the systematic modelling of VDTs. The contributions to practice and the potential for further research based on these findings were then also outlined.

This article is organized as follows: Section 2 first describes the methodological concept and the practical applications of VDTs. Then, the necessity of a systematic notation for VDTs is justified. Section 3 describes the methodology, consisting of the procedure for developing the notation and its evaluation. Section 4 presents the notation with its semantics and syntax. Section 5 describes the demonstration of the notation using two case studies. Section 6 documents the evaluation of the notation and its results. Section 7 summarises the contributions of the notation, outlines the implications for research and practice, and discusses potential limitations.

## 2 Value Driver Trees and Modelling Notations

VDTs are a structured system of indicators that, collectively, reflect the value creation logic of a specific business model. For this purpose, the cause-and-effect relationships between a company's strategic objectives and their underlying performance indicators, the so-called value drivers (VDs), are systematically modelled. In this model, the strategic indicators (e.g., 'profit') are gradually broken down into their underlying financial indicators (e.g., 'revenue' and 'costs'), whose development can in turn be traced back to central VDs (e.g., 'return on advertising spend' and 'overall equipment effectiveness'), which can actually be controlled by the company's management (Rappaport, 1998, 1987; Koller et al., 2020; Matthies, 2024). In this way, VDTs provide management with a transparent basis for analysing and simulating value creation processes and, ultimately, for making effective decisions (Wall and Greiling, 2011).

As indicator-based models, VDTs have certain conceptual characteristics (see Matthies, 2024) The key conceptual elements of a VDT are: (1) nodes, which represent indicators; (2) edges, which illustrate the causal links between indicators; and (3) the hierarchical tree structure, which allows the model to be organised in an orderly manner, allowing intuitive navigation through the value creation logic: (1) The indicators employed in a VDT can be categorised into four distinct groups: (a) target indicators, (b) financial sub-indicators, (c) VDs, and (d) external factors (e.g., interest rates). (2) The relationships between indicators can be modelled in two ways: mathematically (e.g., using mathematical operators and formulas) and logically (e.g., using assumed, non-measurable causal relationships). A further distinction can be made between variables that can be influenced (VDs) and those that cannot (external factors). (3) The conceptual framework underpinning the tree structure is predicated on the target indicator (root node) of the tree structure, which unfolds a network of causal relationships from the financial sub-indicators to the subordinate VDs and external factors (child nodes).

VDTs are increasingly being used in various areas of corporate management, but in particular for planning purposes, in simulations of decisions and developments or in management reporting. However, modern applications – such as predictive forecasting, AI-based planning or automated reporting – require the underlying business models to be available in a structured, complete and, ideally, machine-readable form. VDTs are the appropriate information model for this, as they provide the necessary representation of calculable cause-and-effect relationships of business indicators for such applications.

Despite the increasing importance of VDTs in practice and research, a standardised, well-founded graphical notation for their modelling is still lacking. This is surprising given that VDTs are conceptual models, whose usefulness depends largely on their consistency, completeness and ultimate interpretability. In the field of conceptual modelling – for example in process or data modelling (see e.g., Parsons & Wand, 2008; Recker et al., 2011) – it has long been recognised that systematic visualisation using standardised modelling languages contributes significantly to consistency, completeness and thus interpretability (Moody, 2009; Clark et al., 2015). In the absence of such a notation, as is currently the case with VDTs, modelling is dependent on individual design decisions, which lead to widely varying representations. A review of existing VDTs revealed a wide variance in terms of the conceptual constructs used, the level of detail and the visualisation. Figure 1 gives an impression of the heterogeneity in the model representation based on two exemplary VDTs. These models differ not only in the type of node visualisation (e.g., with or without additional information such as responsibility, data source or development), but also in the representation of relations (e.g., explicit versus implicit edges) and in the structural design (e.g., linear versus modular organisation) (see Matthies, 2024).

**Figure 1**  Exemplary VDT models (Matthies, 2024)

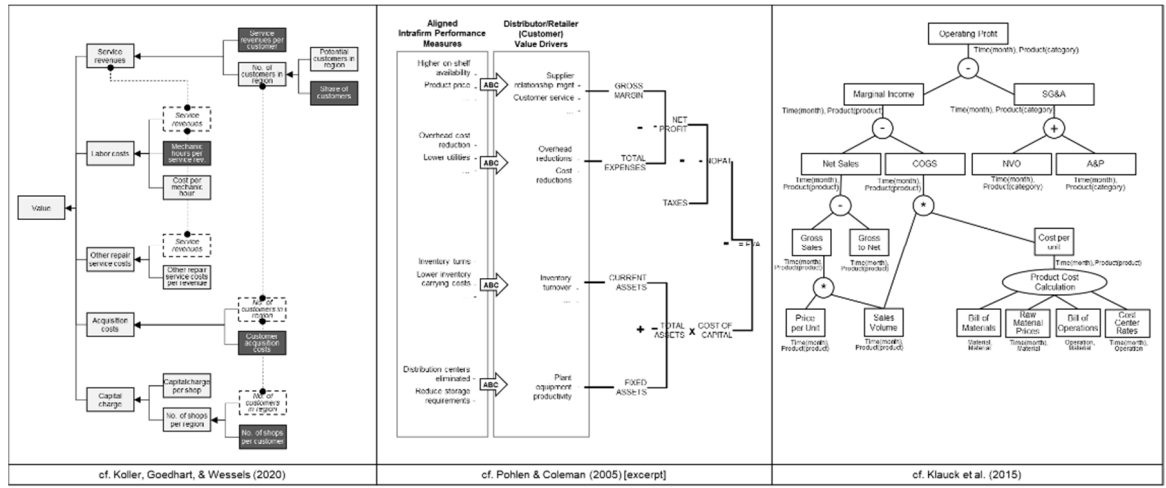

This diversity in VDT modelling is problematic as it increases the cognitive load during interpretation and leaves room for subjectivity, which can lead to incorrect decisions during its modelling and application (Belle et al., 2022). Research into performance management shows that the way indicators are presented can have a big effect on the quality of decisions (Farrell et al., 2012; Kahneman et al., 2016). Therefore, a standardised notation would make a significant contribution to improving the quality of the modelling and application of VDTs.

The development of a graphical modelling notation for VDTs (see Section 3) is based on the extensive research on conceptual modelling (see, e.g., Moody, 2009; Wand & Weber, 2002). A fundamental distinction is made between the semantics and syntax of the notation. In established modelling languages – such as BPMN for business processes or UML for object-oriented systems – this separation is an essential part of the notation design. Semantics (= What is represented?) describes which constructive elements can be represented in a model (e.g., target indicators, financial indicators, VDs, external factors), while syntax (= How is it represented?) deals with their visual representation (e.g., as nodes, lines, symbols), which together result in a uniform 'visual language'.

## 3   Research Design

The design-oriented goal of developing a graphical notation for modelling VDTs follows a Design Science Research (DSR) approach, which aims to design innovative artefacts to solve relevant practical problems (see Hevner et al., 2004; Peffers et al., 2007). This research follows the Design Science Research Methodology (DSRM) proposed by Peffers et al (2007), which consists of the following generic process with six, partially iterative steps (see Figure 2): (1) *Problem identification and motivation*: The relevance of the problem, which is the lack of standardisation in VDT modelling, was confirmed by a comprehensive literature review by Matthies (2024). (2) *Define the objectives for a solution*: The research objective is, as described in Section 1, the development and evaluation of a visual notation for the modelling of VDTs. (3) *Design and development*: Conceptually, the notation is based on a VDT Model Classification developed by Matthies (2024). For this classification, a structured literature review was conducted and 161 published VDTs from research and practice were analysed using an Extended Taxonomy Development Process (Kundisch et al., 2021). As a result, 34 typical conceptual constructs for modelling VDTs were identified and classified. These 34 model constructs cover the semantics of the notation and answer the question 'What is modelled in VDTs?'. Building on this, the syntactic question 'How are the model constructs represented?' is answered in this study by developing a graphical modelling notation. (4) *Demonstration* and (5) *evaluation*: For the practical demonstration and evaluation of the developed notation, this DSR study uses an illustrative scenario approach, which aims at the '[a]pplication of an artifact to a synthetic or real-world situation aimed at illustrating suitability or utility of the artifact' (Peffers et al., 2012, p. 5). In this context, the utility of the developed notation is to be demonstrated in realistic case studies and the fulfilment of central quality criteria for modelling languages is to be evaluated. For this purpose, two existing VDTs, which were developed without a notation, will be transferred to the proposed notation, i.e., they will be 're-modelled'. Care was taken to ensure that the semantic content of the original models remained unchanged, so that a direct comparison of the representations was possible. The reconstructed models were then compared with the original versions and checked for the ability of the new notation to be applied. This was assessed by means of expert evaluations (Peffers et al., 2012) using established criteria for assessing the quality of modelling languages (cf. Moody, 2005). The knowledge gained will contribute to the further improvement of the notation and underpin its relevance in practice. (6) *Communication*: The artefact developed, its potential contributions and further development paths are communicated on the basis of this article (see Section 7).

**Figure 2** Research process

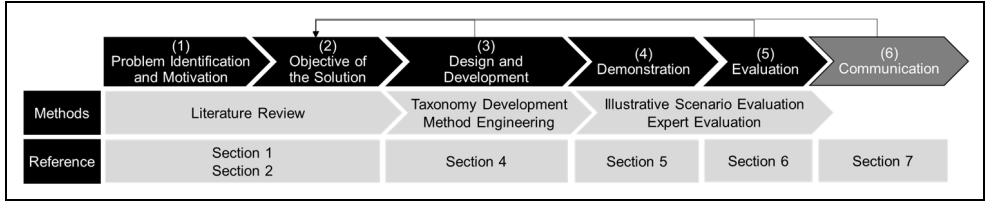

## 4  Value Driver Modelling Notation (VDMN)

The graphical notation for modelling VDTs proposed in this study is called 'Value Driver Modelling Notation' (VDMN; see Figure 3). The modelling notation comprises not only a defined semantics – i.e., a structured collection of typical conceptional constructs such as target indicators, financial sub-indicators, VDs or external factors – but also a graphical syntax based on principles of notation design (see Moody, 2009). This syntax defines, e.g., graphical symbols, connecting lines, operators and ways of presenting additional contextual information such as responsibilities or data sources.

The basic structure of the VDMN (see Figure 3) is based on the VDT Model Classification defined by Matthies (2024), which divides a total of 34 semantic model constructs into three dimensions ([1] indicators, [2] connections, [3] structure) and eight categories (e.g., dimension [2] is structured into the categories [a] links and [b] operators). This classification is intended to reflect the indicator-based nature of VDTs and their hierarchical tree structure. The meaning of the dimensions, categories and semantic constructs of the VDMN is summarised in Table A.1 in the Appendix. Guidance on the basic use of the VDMN is given below along the dimensions and categories:

- **Indicators**: Indicators are the central conceptional elements of a VDT. To ensure a clear, consistent and analytically valuable presentation of indicators, the following recommendations can be made:
  - *Indicator type differentiation*: Indicators should be distinguished according to their inherent characteristics and relevance. Awareness and explicit modelling of these different indicator types can improve the relevance and interpretability of the VDT. In particular, it is advisable to distinguish between (1) key business indicators, (2) financial indicators, (3) VD, (4) external indicators, (5) intermediate results.
  - *Functional highlighting*: Visual highlighting techniques should be used to emphasise the functional role or relevance of certain indicators. This includes, in particular, the use of specific colours or shapes to differentiate, e.g., key VDs, input indicators or calculated (subsidiary) results.
  - *Indicator content enrichment*: The modelling of indicators should not be limited to their basic measurement. Instead, it is advisable to enrich indicators with additional content, such as units of measurement, data sources, comparative benchmarks or graphical elements (e.g., trend arrows). These enrichments facilitate the interpretability of the VDT and support managerial decision-making by providing relevant contextual information.
- **Connections**: The modelling of hierarchical connections in a VDT is crucial to make the causal relationships between indicators transparent and analytically interpretable. Two categories of connections should be considered:
  - *Links*: Logical and analytical relationships between indicators should primarily be visualized through direct links, typically in the form of arrows. These links represent clear cause-effect chains and facilitate a straightforward interpretation of how changes in one indicator affect others. Where appropriate, indirect links may be used to represent cross-hierarchical dependencies or feedback loops; however, their application should be limited to avoid compromising the hierarchical clarity of the tree structure. Although informal groupings of indicators without explicit analytical links are possible, they are generally not recommended as they limit the VDT's explanatory power.

- *Operators*: To specify the mathematical logic underlying the links between indicators, appropriate operators should be included in the model. These include basic mathematical operators (e.g., addition, multiplication) as well as logical conditions (e.g., decision points in the form of "if-then" rules). The explicit modelling of operators enhances the analytical power of the VDT by making computational dependencies transparent. Omitting operators, on the other hand, limits the usability of the model for quantitative analysis and simulation.
- **Structure**: The structural arrangement of a VDT has a significant impact on its clarity and interpretability, and therefore its practical usability. Overly complex or unstructured VDTs can undermine their purpose of reducing complexity and supporting managerial decision-making. The following structural recommendations are therefore proposed:
  - *Levels*: The hierarchical structure of the VDT should be organised into meaningful levels to enhance clarity. These levels can be defined, e.g., by type of indicator (e.g., separation of financial indicators, VDs and external indicators) or by operational functions (e.g., marketing and production). Such structuring supports a more systematic interpretation of the model and highlights the respective responsibilities of different indicators.
  - *Clusters*: Where appropriate, related indicators should be grouped into clusters based on functional, organisational or business model criteria. Clustering facilitates the identification of specific relationships within the model and allows decision makers to focus on relevant subsets of indicators without losing sight of the overall tree structure.
  - *Decompositions*: In order to avoid information overload and to ensure that the VDT remains comprehensible, decomposition techniques can be applied. These include dividing complex VDTs into multiple subtrees or selectively pruning less relevant branches. Previous research suggests that as little as a fourth level of hierarchy can significantly increase complexity and hinder the effective use of VDTs (Matthies, 2024). Therefore, limiting the depth and breadth of VDTs is essential to maintain their analytical usability and managerial relevance.

**Figure 3** VDMN – Value Driver Modelling Notation

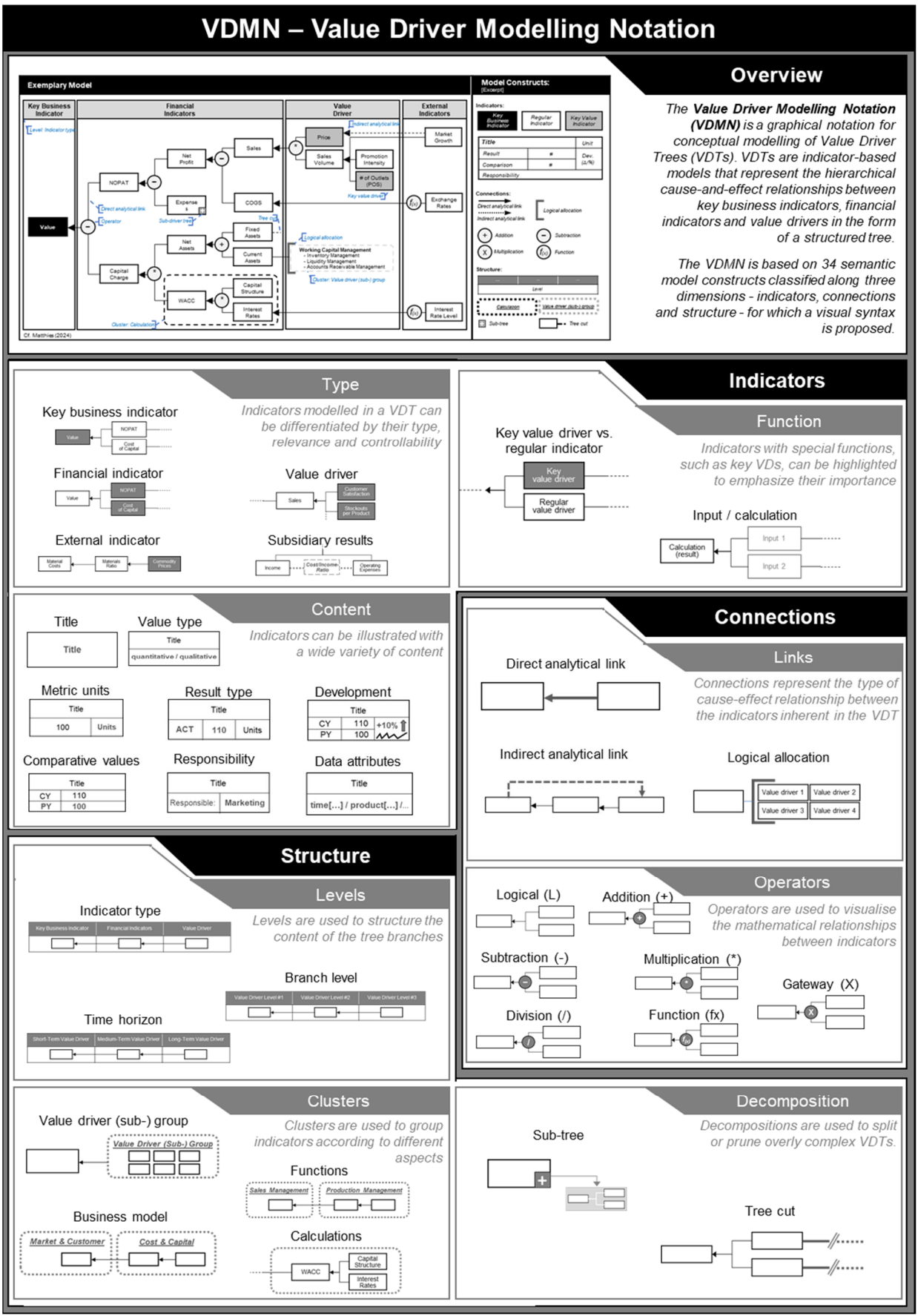

## 5 Demonstration

According to the DSRM (Peffers et al., 2007), the developed artefact must be demonstrated in a practical application and then evaluated. For this purpose, two examples of practical VDTs, initially modelled without a systematic notation, were then 're-modelled' using the VDMN. The aim was not only to demonstrate its application, but also to evaluate the experience gained during the modelling process using systematic criteria (see Section 6).

The first demonstration involved modelling a VDT with the target indicator 'Gross Profit'. Figure 4 shows the model without the use of any notation and with the application of the VDMN. The following modelling constructs (comment boxes mark the constructs in the figure) were used for specific purposes:

- *Indicator type*: The different types of indicators used – key business indicator, financial indicator, VDs and external indicators – have been grouped according to their respective levels (see below).
- *Indicator function*: Key indicators with a central function have been visually highlighted, such as the key business indicator (black), key VDs (grey) and subsidiary (result) indicators (dashed lines), in order to increase focus.
- *Links*: In addition to direct analytical links between indicators, which reflect the logical direction of the hierarchical tree structure, an indirect analytical link was also used, which also represents a mathematical relationship, but provides a connection back into the logically structured, tree-like hierarchy.
- *Operators*: The use of operators facilitates the interpretation of mathematical relationships.
- *Levels*: The VDT is hierarchically organised into levels according to the different types of indicators to facilitate focus.
- *Clusters*: The use of clusters organises VDs according to business functions, which increases the context and defines responsibilities for key VDs.
- *Decomposition*: The use of a tree cut reduces complexity and hides indicators that are irrelevant for the initial interpretation of the VDT.

The second demonstration involved modelling a VDT with the target indicator 'Return on Capital Employed (ROCE)' (see Figure A.1 in the Appendix). The following VDMN modelling constructs were used for specific purposes:

- *Indicator type*: The VDT uses the standard indicator types – key business indicator, financial indicator, VDs, external indicators and a subsidiary (results) indicator.
- *Indicator function*: In this VDT, indicators are visually differentiated specifically according to their function. In addition to the key business indicator (black), the central VDs, subsidiary indicators (dashed), input variables are also specially indexed (transparent), which are less relevant for steering and in particular represent the data for a subsequent calculation.
- *Links*: In addition to direct analytical links, a logical allocation of VDs was modelled to illustrate that VDs (e.g., training and automation) affect a higher-level indicator (COGS) without actually integrating a mathematical relationship.
- *Operators*: The use of operators facilitates the interpretation of mathematical relationships.
- *Levels*: The VDT is hierarchically organised into levels according to the different types of indicators to facilitate focus.
- *Cluster*: Clustering is used to indicate that certain VDs have a logical, but not individually specified (measurable), influence on the assigned indicator (COGS).

## 6 Evaluation

The VDMN is evaluated using established criteria for conceptual modelling grammars (*construct deficit*, *construct redundancy*, *construct excess*, *construct overload*; adapted from Wand &Weber, 1993, 2002), which should reveal potential deficiencies of the modelling notation. To evaluate these criteria, the VDTs reconstructed using the VDMN were compared with the original models to check for semantic and syntactic congruence. Feedback interviews were then conducted with the two experts (#1-2) to discuss the reconstructed models and gain their insights into the models' informational value and the utility of the VDMN for modelling purposes.

- *Construct deficit*: According to this criterion, it is evaluated whether the notation does not contain all the conceptional constructs required for comprehensive modelling of VDTs. If any concepts were missing, the notation would be incomplete. The corresponding evaluation revealed no indications of a construct deficit in the VDMN. All conceptual constructs required for comprehensive modelling of the demonstrative VDTs were available. Experts emphasised that the notation allows even complex aspects to be modelled, like contextual aspects from the company: 'With levels or clusters, it is possible to flexibly integrate even two types of contextual aspects, such as the structuring of indicators and, at the same time, their allocation to company functions' (expert #2).

- *Construct redundancy*: According to this criterion, it is evaluated whether the notation provides more than one explicit conceptual construct to represent the same specific element (ontological construct) of a VDT model. The corresponding evaluation revealed no real evidence of construct redundancy within the VDMN. Nevertheless, two aspects were highlighted that require explicit guidance for modelling. First, both levels and clusters are considered to be structuring constructs. Despite the apparent benefits of both constructs, particularly the combination of these, further guidance is required to determine the most suitable application of each for specific structuring purposes (e.g., levels for structuring indicators and clusters for corporate responsibilities). Second, it was highlighted that two types of VDT decomposition also require distinct instructions for individual use (e.g., sub-trees for the subdivision into independent VDTs and tree cuts for the shortening of individual indicators): 'Although both sub-tress and tree cuts are useful, it would be helpful to know when and how to use each one specifically' (expert #1).
- *Construct excess*: According to this criterion, it is evaluated whether the notation contains conceptual constructs that are meaningless for the purpose of modelling VDTs. The corresponding evaluation confirms that the VDMN contains no excessive constructs that are irrelevant for the modelling of VDTs: 'There seems to be no item [construct] that just takes up space without adding any informative value' (expert #2). However, it must be acknowledged that the evaluation carried out so far can only be considered as preliminary in this respect. As certain constructs of the VDMN have not yet been used in the two demonstrative VDTs, a possible excess of constructs cannot be ruled out entirely.
- *Construct overload*: According to this criterion, it is evaluated whether the notation contains conceptual constructs that have more than one specific meaning for VDT modelling, meaning they could be used for more than one modelling purpose. The corresponding evaluation revealed no signs of construct overload in the VDMN. Each conceptual construct was associated with a single, unambiguous meaning, avoiding semantic confusion. In this respect, a specific recommendation for modelling was developed: In order to differentiate between the different indicator types, which are basically modelled using rectangles, the structuring use of levels was considered particularly useful, which excludes the potential ambiguity of indicators.

In addition to discussing the above criteria, qualitative assessments of the *interpretability* of the modelled VDTs (as suggested by Moody, 2005) were also collected. The standardised representation of VDTs not only facilitates the comparison of different models, but also facilitates intuitive comprehension. In particular, the structuring elements of the notation allow for the integration of contextual information, such as the differentiation of indicator types, responsibilities or relationships between calculations. In addition, the integration of mathematical operators is considered essential. Altogether, this explicit contextualisation of indicators facilitates the interpretation of the VDT statements. This is in line with findings that the way indicators and their contextual relationships are presented significantly influences managers' ability to interpret them accurately and make effective decisions (see, e.g., Belle et al., 2022; Farrell et al., 2012). In this respect, the targeted and differentiated use of nuanced visual aids, such as highlighting indicators with a special function (e.g., key VD), is also considered helpful.

**Figure 4** Demonstration 1 'Gross Profit'

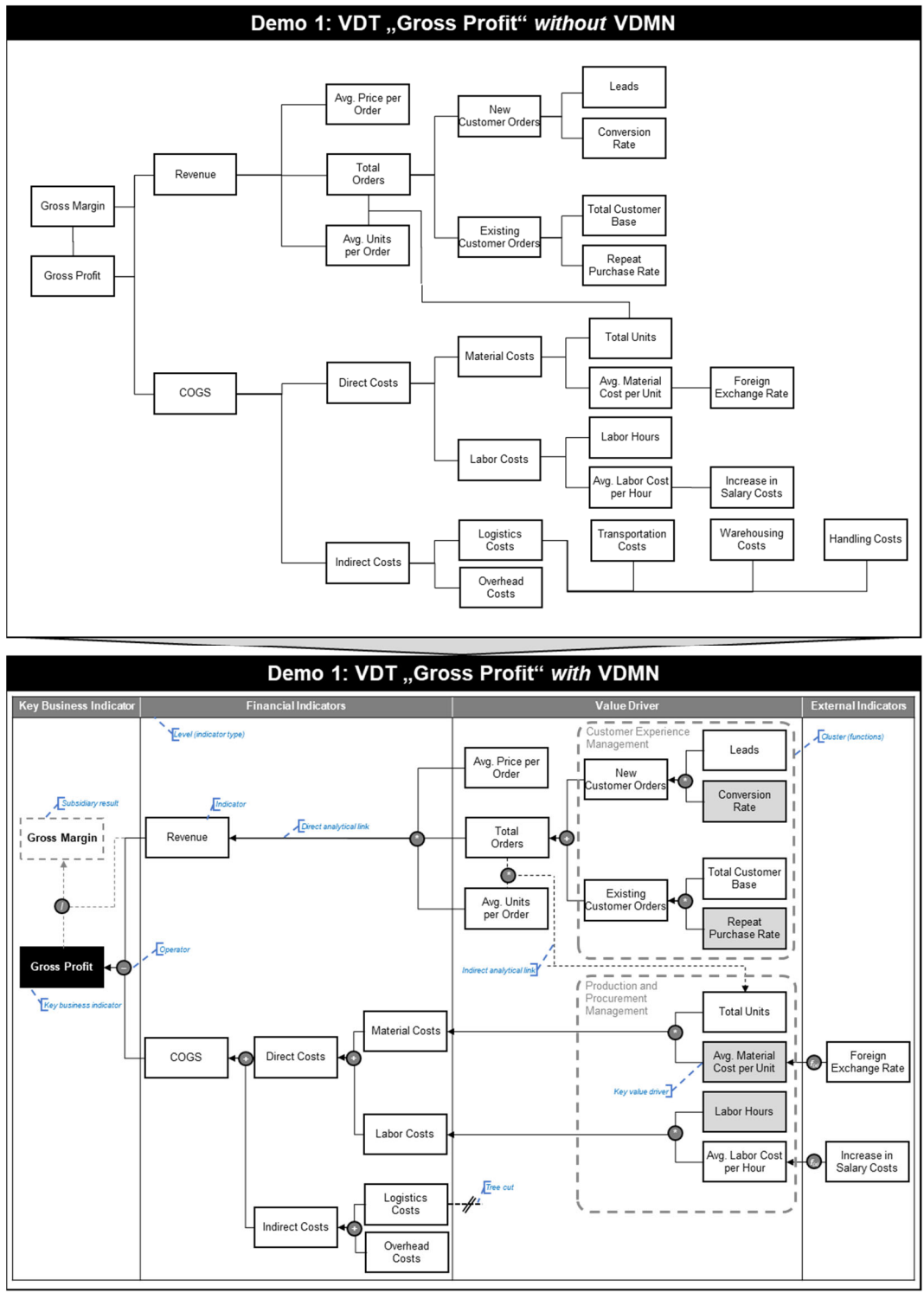

## 7 Conclusions

The VDMN provides a consistent set of conceptual constructs and syntactic tools that enable the visualisation and interpretation of complex cause-and-effect relationships between strategic, financial and operational indicators in VDTs. As such, the developed notation is a significant step towards substantiating and standardising VDT modelling. Better developed VDTs would not only improve their use by management but also significantly enhance their connectivity to modern technologies, such as AI-based planning systems. Thereby, the notation presented here can therefore lay the conceptual foundations for future data-driven management models.

The VDMN addresses several challenges of current practice in VDT modelling. First, it contributes to methodological standardisation, thereby reducing the conceptual heterogeneity of VDT models and replacing it with uniform semantic constructs and syntactic rules. Second, it substantiates and improves the communicative function of VDTs by making the models more interpretable and comprehensible in terms of informational value for both VDT modellers and decision makers. Therefore, practitioners are advised to adopt the VDMN when developing or revising VDTs, in order to ensure consistency, enhance interpretability, and improve informational value. To promote this practical application, it is planned to publish the notation, including case studies and application instructions, on an individual website.

From a research perspective, the VDMN provides a conceptual foundation for further investigation into VDT modelling. First, future research should focus on a more extensive evaluation of the VDMN in diverse organisational contexts, including quantitative assessments of model quality and modelling performance. After evaluating the consistency and completeness of the notation, the next step is to further investigate usage and interpretation performance of VDT users. The evaluation criteria proposed by Paige et al. (2000) and Burton-Jones et al. (2009) can be useful for this purpose (e.g., interpretational fidelity and efficiency). Second, also building on the experience gained in future evaluations, further research should focus on the development of additional, more specific modelling guidelines. The evaluation conducted in this paper has already identified specific needs (e.g., individual application guidance for sub-trees and tree cuts). Finally, the development of software tools that support the systematic modelling and validation of VDMN-based models represents a promising direction for applied research.

This study also contains certain limitations. First, the development of the VDMN is based on an ex-post classification of existing VDTs, which, although comprehensive, may still reflect inherent biases in the source material. Second, the empirical evaluation of the notation was limited to two case studies and expert interviews. While this allowed for a qualitative assessment of the usability and consistency of the notation, the generalisability of the findings remains limited. Further evaluations with larger, more diverse cases are therefore needed to consolidate the findings and fully validate the practical utility of the VDMN.

# Appendix

**Table A.1** VDMN Constructs (cf. Matthies, 2024, `VDT Modelling Classification´) [part 1/3]

| Dimension | Category | Construct | Description |
|---|---|---|---|
| | Indicators | | Indicators are the key elements to be modelled in VDTs, whose meaning and representation can be characterized by three categories: type, function and content. |
| | | Type | Differentiation of the indicators modelled in a VDT in terms of type, relevance and controllability. |
| | | Key Business Indicator | Strategic target indicator at the top of the VDT representing a value of central relevance to the business, whose result is explained by the indicators below. |
| | | Financial Indicator | Monetary indicator that represents a financial (interim) result that is influenced by the business performance (value driver) of the company. |
| | | Value Driver | Indicators that represent certain factors of business performance that can be systematically influenced by the company and that has a significant impact on the higher-level financial indicators and, ultimately, the key business indicator. |
| | | External Indicator | Factors in a company's external environment that affect its performance but cannot be actively controlled by the company. |
| | | Subsidiary Results | Secondary results that do not play a direct role in the logical hierarchy of the VDT, but are associated with it and are of interest as interim indicators. |
| | Function | | Indicators modelled in a VDT can be differentiated by their type, relevance and controllability. |
| | | Key Value Indicator/ Regular Indicator | Value drivers that are particularly relevant (because of their influence on the key business indicator) and should therefore be visually highlighted in the presentation of the VDT. |
| | | Input/ Calculation | Visual distinction between the manageable input of a calculation and the derived (unmanageable) results. |
| | Content | | Indicators can be illustrated with a wide variety of content. |
| | | Title | Meaningful name of the indicator. |
| | | Value Type | Additional specification of the value type (e.g., quantitative vs. qualitative, leading vs. lagging). |
| | | Metric Unit | Units of quantitative measurement (e.g., $, pieces, $/piece, %). |
| | | Data Attributes | Data management attributes (e.g., table, time, product, category). |
| | | Result Type | Type of result (e.g., actual, budget, forecast). |
| | | Comparative Vales | Specification of a second comparison value (e.g., current year vs. previous year). |
| | | Development | Visual or quantitative indication of trends. |
| | | Responsibility | Organisational unit or person responsible for the indicator. |

**Table A.1** VDMN Constructs (cf. Matthies, 2024, `VDT Modelling Classification´) [part 2/3]

| Dimension | Category | Construct | Description |
|---|---|---|---|
| Connections | | | Connections represent the type of cause-effect relationship between the indicators inherent in the VDT. |
| | Links | | Links connect indicators according to their relationship (e.g., mathematically). |
| | | Direct Analytical Link | Direct and immediate analytic relationship between indicators in the VDT hierarchy. |
| | | Indirect Analytical Link | Secondary and intermediate relationships between indicators in the VDT hierarchy that are analytically evident. |
| | | Logical Allocation | Logical assignment of an indicator to a higher-level indicator that cannot be modelled explicitly in an analytical way. |
| | Operators | | Operators are used to visualise the mathematical relationships between indicators, thereby simplifying the interpretation of cause-effect chains. |
| | | Logical (L) | Logical relationship between indicators, without further indication of its calculation. |
| | | Addition (+) | Addition of indicators (child nodes) to a result (parent node) following in the VDT hierarchy. |
| | | Substraction (-) | Subtraction of indicators (child nodes) to a result (parent node) following in the VDT hierarchy. |
| | | Muliplication (*) | Multiplication of indicators (child nodes) to a result (parent node) following in the VDT hierarchy. |
| | | Division (:) | Division of indicators (child nodes) to a result (parent node) following in the VDT hierarchy. |
| | | Function ($f_x$) | Mathematical function (e.g., calculation of averages or the application of a statistical regression function) that processes the input of indicators (child nodes) to produce a result (parent node) that follows in the VDT hierarchy. |
| | | Gateway (X) | Decision point in a VDT hierarchy that can change the path of the VDT (e.g., either path A or path B) under certain conditions (e.g., the value of a previous indicator). |

**Table A.1** VDMN Constructs (cf. Matthies, 2024, `VDT Modelling Classification´) [part 3/3]

| Structure | | | Structure-creating elements are used to organise the indicators of comprehensive VDTs for interpretation. |
|---|---|---|---|
| | Levels | | Levels are used to structure the content of the tree branches. |
| | | Indicator Type | Structure of the VDT hierarchy by type of modelled indicators. |
| | | Branch Level | Structure of the VDT according to the number of hierarchical levels. |
| | | Time Horizon | Structure of the VDT hierarchy according to management timeframes. |
| | Clusters | | Clusters are used to group indicators according to different aspects |
| | | Value Driver (Sub-) Group | Clustering of VDs that have a logical, but not individually specified, influence on an assigned indicator. |
| | | Business Model | Clustering of indicators based on their association with specific areas of the business model. |
| | | Functions | Clustering of indicators based on their association with specific functional or organisational areas of the company. |
| | | Calculation | Clustering of indicators based on their association with a calculation to be specifically displayed on the VDT. |
| | Decomposition | | Decompositions are used to split or prune overly complex VDTs. |
| | | Sub-Tree | Decomposition of a VDT by representing more extensive connections of an indicator in a separate sub-VDT. |
| | | Tree Cut | Pruning of a VDT by neglecting more extensive connections and merely referencing them. |

**Figure A.1**    Demonstration 2 'Return on Capital Employed (ROCE)'

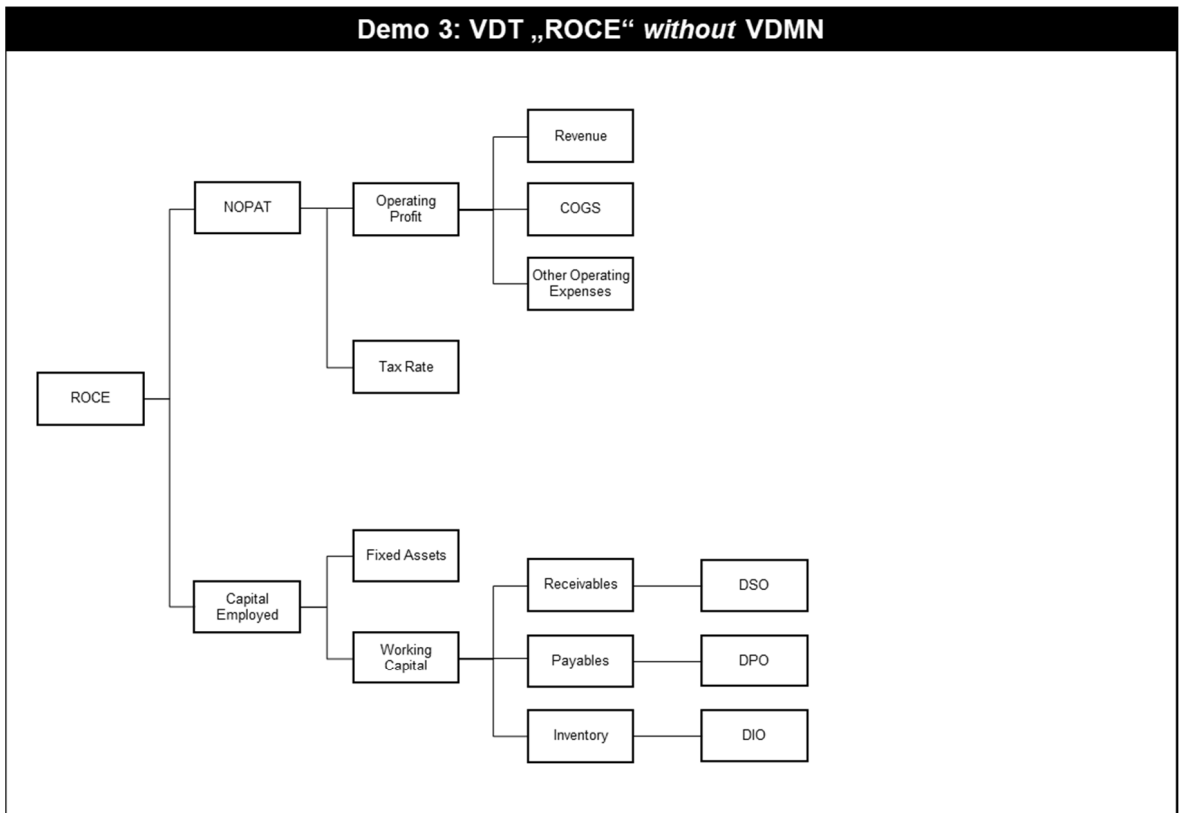

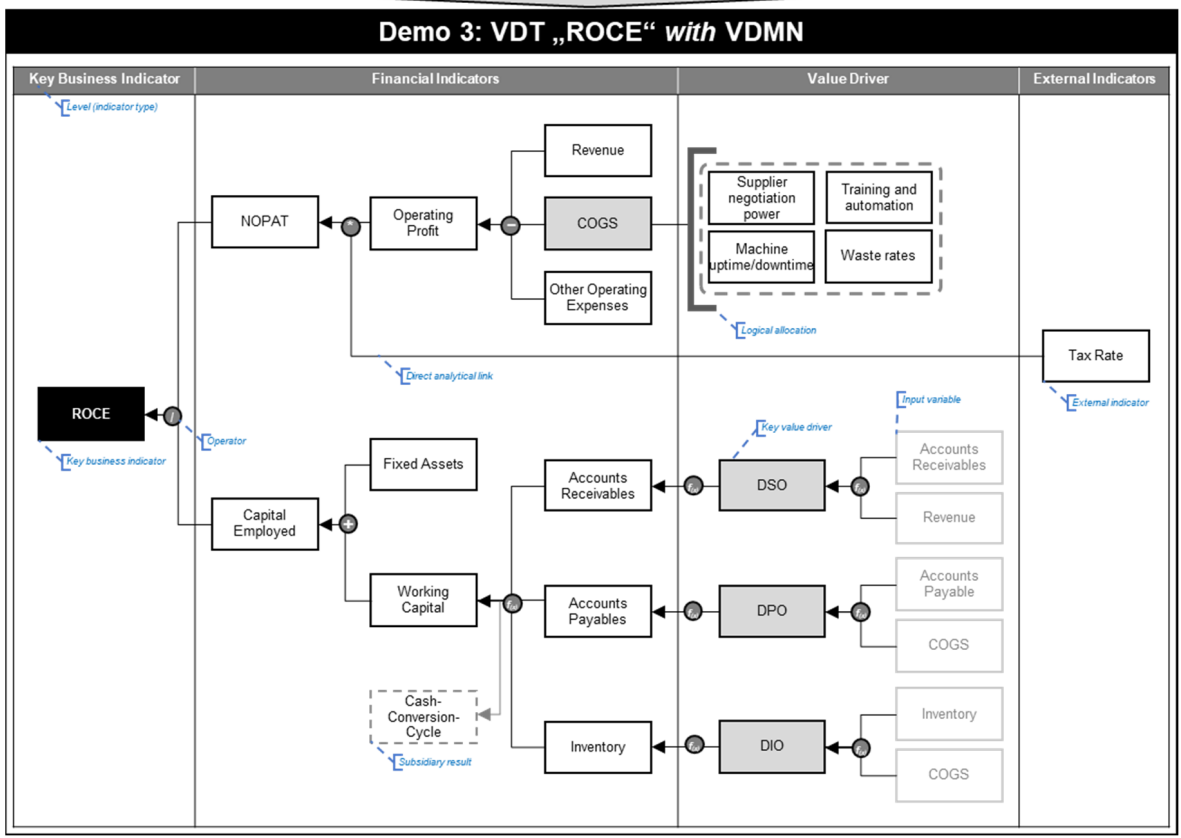